\documentstyle[float,twocolumn,prl,aps,epsfig]{revtex} 
\newcommand{\bea}{\begin{eqnarray}}
\newcommand{\eea}{\end{eqnarray}}
\newcommand{\gapp}{\raisebox{-.5ex}{$\stackrel{>}{\sim}$}}

\newcommand{\la}{\langle}
\newcommand{\ra}{\rangle}

\begin{document}
\preprint{\\CERN-TH/2001-???\\LBNL-47405\\hep-ph/0101136} 
\title{Radial and Elliptic Flow at RHIC: Further Predictions}

\author{P.~Huovinen$^a$, P.F.~Kolb$^{b,c}$, U.~Heinz$^b$, 
        P.V.~Ruuskanen$^d$, and S.A. Voloshin$^e$}
\address{$^a$Lawrence Berkeley National Laboratory, Berkeley CA 94720, USA\\
         $^b$Department of Physics, The Ohio State University, 174 West
         18th Avenue, Columbus, OH 43210, USA\\
         $^c$Institut f\"ur Theoretische Physik, Universit\"at
         Regensburg, D-93040 Regensburg, Germany\\
         $^d$Department of Physics, University of Jyv\"askyl\"a, FIN-40341
         Jyv\"askyl\"a, Finland\\
         $^e$Department of Physics and Astronomy, Wayne State University,
         666 W. Hancock Street, Detroit, MI 48202, USA}
\date{\today}

\maketitle

\begin{abstract}
Using a hydrodynamic model, we predict the transverse momentum dependence 
of the spectra and the elliptic flow for different hadrons in Au+Au 
collisions at $\sqrt{s}=130\,A$\,GeV. The dependence of the differential 
and $p_t$-integrated elliptic flow on the hadron mass, equation of 
state and freeze-out temperature is studied both numerically and 
analytically.
\end{abstract}

\smallskip



\medskip


{\it 1. Introduction.--} 
One of the first observables measured at the Relativistic Heavy Ion 
Collider (RHIC) was the so-called elliptic flow \cite{STAR00}. It 
describes the azimuthal momentum space anisotropy of particle emission 
from non-central heavy-ion collisions in the plane transverse to the 
beam direction. Elliptic flow is characterized by the second harmonic 
coefficient $v_2({\rm y},p_t)$ of an azimuthal Fourier decomposition of 
the momentum distribution \cite{O92,VZ96}. We here discuss elliptic flow 
at midrapidity, ${\rm y}=0$; its $p_t$-averaged value is denoted simply 
by $v_2$. 

Elliptic flow is a fundamental observable since it directly reflects the
rescattering among the produced particles. Rescattering transfers the 
initial spatial anisotropy of the nuclear overlap region in the transverse
plane to the observed momentum distribution. For a given initial spatial
deformation, the largest elliptic flow coefficient is obtained in the
hydrodynamic limit where rescattering is so intense that the matter in
the reaction zone reaches a state of local thermal equilibrium. Since
the spatial anisotropy is largest at the beginning of the evolution,
elliptic flow is especially sensitive to the early stages of system
evolution \cite{S97,KSH99}. A measurement of $v_2$ thus provides access
to the fundamental thermalization time scale in the early stages of
a relativistic heavy-ion collision \cite{VP00,KSH00}.

In a preceding Letter we showed \cite{KH3} that the first measurements 
at RHIC of the elliptic flow of charged particles can be satisfactorily 
described by a hydrodynamical model, with initial and freeze-out 
conditions obtained by a straightforward extrapolation from a similar
analysis of Pb+Pb collisions at the SPS \cite{KSH99,KSH00}. The data 
deviate, however, from the hydrodynamical prediction at large impact 
parameters $b\gapp 7$\,fm, where the reaction volumes become small,
and at large transverse momenta $p_t\gapp 1.5$\,GeV/$c$. At the lower 
SPS energies ($\sqrt{s}=17\,A$\,GeV) the available data 
\cite{NA49v2,NA49QM99} are less conclusive, but indicate discrepancies 
with the hydrodynamic approach for all but nearly central collisions 
\cite{KH3,TLS00}. Whereas in \cite{STAR00,KH3} the deviations from 
hydrodynamic predictions were interpreted to signal incomplete 
thermalization, the authors of \cite{TLS00} suggest (at least in 
connection with the SPS data) that they may be due to the transition 
from an early hydrodynamic to a late hadronic kinetic stage, modeled 
by a hadronic cascade (RQMD) which leads to an earlier saturation of 
both radial and elliptic flow. 

In the present paper we present hydrodynamic predictions at 
$\sqrt{s}=130\,A$\,GeV for the shape and impact parameter dependence of 
the single particle spectra and for the magnitude and $p_t$-dependence 
of the elliptic flow for a variety of hadron species. Such data should 
soon become available. They will not only provide evidence for the degree 
of thermalization of other hadronic species and thus provide valuable 
input for the discussion of the two different interpretations just 
mentioned \cite{KH3,TLS00}, but also help to select between the 
different combinations of initial and freeze-out conditions and 
equations of state studied in \cite{KH3}. 

{\it 2. Hydrodynamic Results.--}
The hydrodynamic code, its initialization and the calculation of 
final state spectra and elliptic flow coefficients are described in 
\cite{KSH00,KH3}. We here use the parameter sets given in the last 
three columns of Table I in \cite{KH3} and label the results 
accordingly by EOS~Q(120), EOS~Q(140), and EOS~H(140), respectively.
EOS~H is a hadron resonance gas equation of state with sound velocity
$c_{\rm s}\approx \sqrt{0.15}$; EOS~Q includes a first order phase 
transition to an ideal quark-gluon gas ($c_{\rm s}=\sqrt{1/3}$) at 
$T_{\rm c}=164$\,MeV with a latent heat of 1.15 GeV/fm$^3$ 
\cite{SHKRPV}. In the energy region studied here, EOS~Q is effectively 
softer than EOS~H, since the expansion is largely controlled by the
soft phase-transition region \cite{KSH00}. Freeze-out is implemented 
along a surface of constant energy density; the numbers in brackets 
denote the approximate freeze-out temperature in MeV on this surface.  

\begin{figure}
   \hspace*{0.1cm} 
   \epsfxsize 7cm \epsfbox{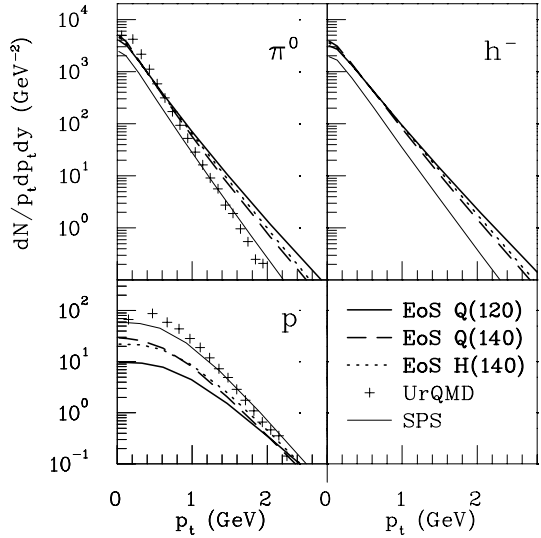}
\caption{The $p_t$ spectra of neutral pions (upper left), negative 
         hadrons (upper right), and protons (lower left) for the 6$\%$ 
         most central Au+Au collisions at $\sqrt{s}=130\,A$\,GeV, for 
         different equations of state and freeze-out temperatures (see 
         text). The crosses show UrQMD results for neutral pions and 
         protons at $\sqrt{s}=200\,A$\,GeV \protect\cite{Bleicher99}  
         for comparison. Also shown are the corresponding hydrodynamic 
         spectra at $\sqrt{s}=17\,A$\,GeV for EOS~Q(120) (thin solid 
         lines).
}
\label{F1}
\end{figure}

{\it Radial flow and single-particle spectra:} 
The transverse momentum spectra for negative hadrons, neutral pions and
protons are shown in Figure~\ref{F1}. All decay products from strong or
electromagnetic decays of unstable resonances up to 1.4~GeV mass are included.
These $p_t$-spectra are considerably flatter than the corresponding SPS 
spectra; the thin solid lines show our earlier fits \cite{KSH99,KSH00} to 
the data from Pb+Pb collisions at $\sqrt{s}=17\,A$\,GeV \cite{NA49spectra}.
This reflects the stronger radial flow at RHIC which hydrodynamics 
predicts as a result of the higher initial energy density \cite{KSH00}. 
The crosses in Figure~\ref{F1} show that the hadronic cascade code UrQMD 
\cite{UrQMD} gives much steeper spectra \cite{Bleicher99}; 
in fact, the UrQMD spectra at RHIC energies are slightly steeper than the 
measured spectra at SPS energies. Apparently the scattering mechanisms 
built into UrQMD are not efficient enough to build a sufficient amount 
of radial flow. On the other hand, the flatter slopes of hydrodynamic 
pion spectra are consistent with first preliminary data on negative
hadron spectra at midrapidity \cite{STARprel}. The final data are 
expected to be accurate enough to distinguish between the three 
parameter sets shown in Figure~\ref{F1} (or exclude all of them).

\begin{figure}
  \begin{center}
   \hspace*{-0.4cm} \epsfxsize 7cm 
     \epsfbox{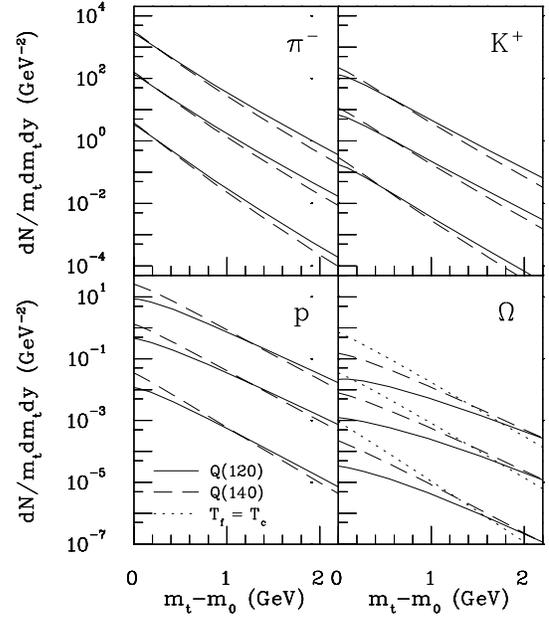}
  \end{center} 
\caption{The $m_t$-spectra of negative pions (upper left), kaons
        (upper right), protons (lower left) and $\Omega$ baryons (lower right) 
         for Au+Au collisions at $\sqrt{s}=130\,A$\,GeV, for collision
         centralities (top to bottom) $b < 5.4$, $5.4 < b < 9.9$ and
         $9.9 < b < 13.5$~fm. For clarity the spectra for
         different centrality bins are separated by factors of 10. The
         calculations were done with EOS Q. The $\Omega$ distribution is
         also shown for $T_f = 164$~MeV to simulate decoupling at the 
         phase transition.}
\label{F2}
\end{figure}

Figure~\ref{F2} shows $m_t$-spectra for pions, kaons, protons and $\Omega$ 
hyperons at different collision centralities. For peripheral collisions 
the spectra get steeper, due to a decrease of the average radial flow 
velocity $\la v_\perp \ra$ as a result of a lower initial energy density 
and a shorter lifetime of the reaction zone. The comparison between theory 
and experiment will be particularly interesting in the region of large 
impact parameters where the measured elliptic flow \cite{STAR00} lags 
behind the hydrodynamical results \cite{KH3}. As discussed in the 
Introduction, the single-particle spectra may help to understand whether 
the origin of this discrepancy is incomplete thermalization at an early 
or late stage of the evolution.
 
\begin{center}
\begin{tabular}{|l|c|c|c|}  \hline
 EOS          &    Q    &   Q    &   H  \\ \hline
$T_{\rm f}$ (MeV) $\approx$ 
              &   120   &  140   &  140 \\ \hline
$\pi^-$       & 4.4\%   &  3.6\% & 4.2\% \\
$K^+$         & 5.3\%   &  4.7\% & 5.7\% \\ 
$p$           & 5.9\%   &  5.3\% & 6.6\% \\ 
$\phi$        & 6.2\%   &  5.6\% & 6.9\% \\
$\phi (T_{\rm f}{=}164\,{\rm MeV})$ & 3.8\%  &  3.8\% &  -    \\ 
$\Lambda$     & 6.1\%   &  5.6\% & 7.0\% \\
$\Omega$      & 6.5\%   &  6.2\% & 7.7\% \\
$\Omega (T_{\rm f}{=}164\,{\rm MeV})$ & 4.4\%  &  4.4\% &  -    \\ \hline
\end{tabular} \vspace*{2ex}
\end{center}
{\small
Table 1. Elliptic flow coefficient $v_2$ at midrapidity for various 
        hadron species from minimum bias Au+Au collisions at 
        $\sqrt{s}=130\,A$\,GeV. The $p_t$-average was taken over
        the full spectrum.
}
\vspace{5mm}

{\it Elliptic flow.} The impact parameter and $p_t$ dependence of the
elliptic flow coefficient $v_2$ for charged particles from Au+Au 
collisions at $\sqrt{s}=130\,A$\,GeV was discussed in \cite{KH3}. 
Table~1 shows predictions for the $p_t$-integrated elliptic flow 
of a variety of identified hadrons in minimum bias collisions, for the 
same three parameter sets as studied in that paper. For the $\phi$ meson 
and the $\Omega$ hyperon we include two values, one for simultaneous 
freeze-out with all other hadrons, the other for freeze-out directly 
after hadronization at $T_{\rm c}=164$\,MeV. The latter option accounts 
for the expectation that the heavy and weakly coupled $\Omega$ is not 
likely to participate after hadronization in any modifications of the 
previously established flow pattern~\cite{Omega}. Similar arguments hold
for the $\phi$ meson. Even though at RHIC energies a large fraction 
of the flow is already established before hadronization \cite{KSH00}, 
the additional flow generated afterwards by hadronic rescattering is seen 
to affect the $\Omega$ quite strongly, for both $v_2$ (Table~1) and the 
single-particle slope (Figure~\ref{F2}).

For identified pions the dependence of $v_2$ on EOS and $T_{\rm f}$ is 
similar as for the sum of all charged hadrons (see \cite{KH3}): Lower 
freeze-out temperatures and harder EOS lead to flatter single particle 
spectra and thereby to larger $p_t$-integrated elliptic flow. For 
identified hadrons we see that, as their mass increases, there is
stronger sensitivity to the EoS than that to $T_{\rm f}$ (see also
Figure~\ref{F4} below).

\begin{figure}
   \hspace*{-0.2cm} \epsfxsize 7cm \epsfbox{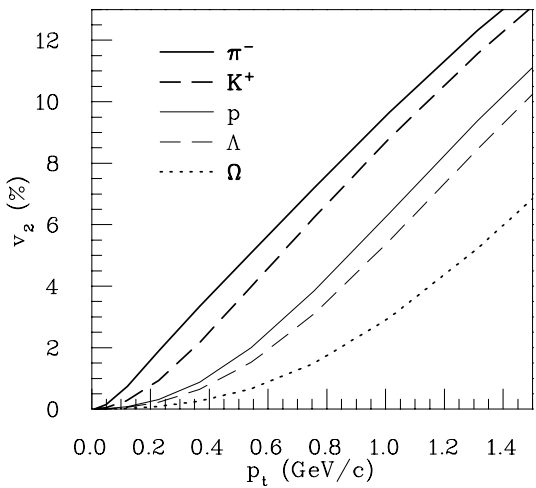}
\caption{$p_t$-differential elliptic flow at midrapidity for va\-ri\-ous 
         hadrons from minimum bias Au+Au collisions at 
         $\sqrt{s}=$ $130\,A$\,GeV for EOS~Q(120).}
\label{F3}
\end{figure}

Figure~\ref{F3} shows the differential momentum anisotropy $v_2(p_t)$ for 
different hadron species for EOS Q and $T_f\approx 120$\,MeV. At a given
value of $p_t$, the elliptic flow is seen to decrease with increasing 
particle mass. This is a consequence of rest-mass-dependent radial flow 
effects on the shape of the single-particle $p_t$-spectrum, as will be 
analytically discussed in the following section.

The smaller differential anisotropy at fixed $p_t$ does not contradict
the results in Table~1 which generically give larger $p_t$-averaged
elliptic flow for heavier particles. This is a consequence of the fact
that {\em radial} flow leads to a flattening of the $p_t$-spectra of
heavier particles \cite{SR79,LHS90}, whose $p_t$-averaged $v_2$ thus
receives more weight from the high-$p_t$ region where $v_2(p_t)$ is
larger. Whether this larger spectral weight for high $p_t$ wins over
the reduction of $v_2$ at fixed $p_t$ depends on the details of the
expansion dynamics.

The effect of the EOS and the freeze-out temperature on the differential
elliptic flow of pions and protons is demonstrated in Figure~\ref{F4}. 
The EOS affects $v_2(p_t)$ for all hadrons in the same way: the stiffer 
EOS~H leads to larger $v_2$ at low $p_t$ and to smaller $v_2$ at high 
$p_t$ than the softer EOS~Q. The effect of the freeze-out temperature 
on $v_2(p_t)$ is more delicate: for pions the effect is small, and for 
EOS~Q an increase in the freeze-out temperature decreases both the 
$p_t$-averaged and the differential elliptic flow. The heavier protons 
behave oppositely: the differential anisotropy increases with increasing
freeze-out temperature. The origin of this behaviour will be studied in 
the following section. Clearly, the different $T_{\rm f}$-dependence of 
$v_2(p_t)$ of different particles can be used to constrain the freeze-out
temperature independently of the EOS.

\begin{figure}
  \begin{center}
   \hspace*{-0.4cm} \epsfxsize 7cm
      \epsfbox{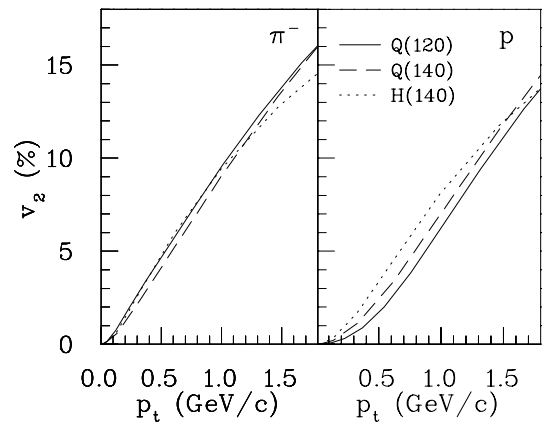}
  \end{center}
\caption{The effect of the EOS and the freeze-out temperature on the elliptic
         flow of midrapidity pions (left) and protons (right) from minimum
         bias Au+Au collisions at $\sqrt{s}=130\,A$\,GeV.}
\label{F4}
\end{figure}

\begin{figure} [ht]
  \begin{center}
   \hspace*{-0.4cm} \epsfxsize 7cm \epsfbox{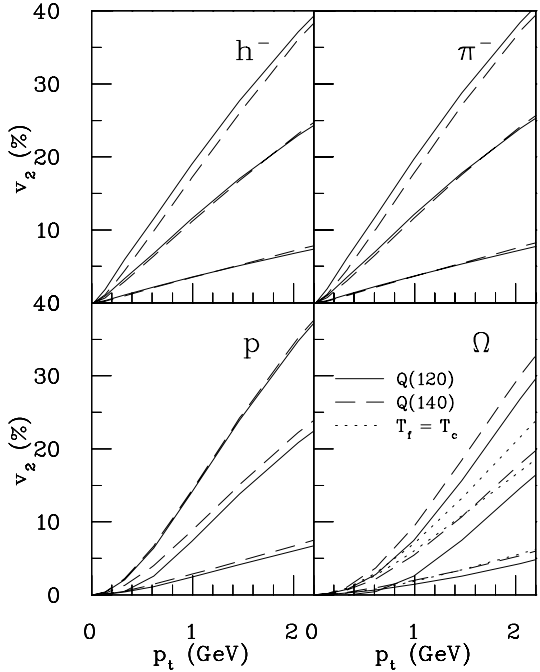}
  \end{center}
\caption{The $p_t$-differential elliptic flow of negative hadrons (upper 
         left), pions (upper right), protons (lower left) and omega baryons
         (lower right) for Au+Au collisions at $\sqrt{s}=130\,A$\,GeV for
         collision centralities (top to bottom) $9.9 < b < 13.5$,
         $5.4 < b < 9.9$ and $b < 5.4$~fm. The calculations were done
         for EOS Q.}
\label{F4-2}
\end{figure}

In view of the observed deviations from hydrodynamic behaviour of
charged particle elliptic flow at large impact parameters \cite{STAR00} 
it will be interesting to study the centrality dependence of $v_2$ 
separately for a variety of hadron species. Corresponding hydrodynamic
predictions are shown in Figure~\ref{F4-2} (again including the option
that the $\Omega$ freezes out early at $T_{\rm f}=T_{\rm c}$). 
Particle-specific deviations from these predictions should provide 
valuable insights into the thermalization and freeze-out mechanisms.

{\it 3. Analytical results.--}
In the remainder we try to understand the hydrodynamic behaviour of 
$v_2(p_t)$ and its dependence on the hadron mass and freeze-out 
temperature, using a simple analytical model. Before going into the 
technical details we give a simple intuitive argument why, at small 
$p_t$, the elliptic flow of heavier particles is smaller than for 
lighter ones. It is well-known that radial flow shifts the 
$p_t$-distributions to larger values of $p_t$, and that for 
nonrelativistic $p_t <  m$ this effect increases with the particle 
mass $m$ and the radial flow velocity $\la v_\perp \ra$. In the extreme 
case of a thin shell expanding at high velocity, the spectrum actually 
develops a relative minimum at $p_t=0$ and a peak at nonzero $p_t$ 
(``blast wave peak'' \cite{SR79}), and with increasing mass and 
$\la v_\perp \ra$ the peak shifts to larger $p_t$. Relative to the 
case without radial flow, the spectrum is thus depleted at small 
$p_t$, and the depletion as well as the $p_t$ range over which it 
occurs increase with $m$ and $\la v_\perp \ra$. 

If, as in the case of fully developed elliptic flow, the radial velocity 
is larger in $x$ than in $y$ direction, $|v_x| > |v_y|$, the same is 
true for this relative depletion effect. It counteracts the excess
of particles with $p_t$ in $x$-direction over those with $p_t$ in 
$y$-direction which is the origin of $v_2>0$. Thus it reduces $v_2$ at
small $p_t$. This reduction and the $p_t$-range over which it occurs
both increase with the particle mass $m$ and the average radial flow
$\la v_\perp \ra$. A quick graphic sketch shows that in the extreme 
case, where the single-particle spectrum develops a ``blast wave peak'' 
\cite{SR79}, $v_2$ even turns negative at low $p_t$, due to the shift
of the peak to larger $p_t$ in $x$ than in $y$ direction if $|v_x| > 
|v_y|$.

When looking for a simple model which allows to show this analytically,
a generalization of the ``blast wave'' model \cite{SR79} comes to mind,
in which thermalized matter of temperature $T_{\rm f}$, approximated by
a boosted Boltzmann distribution, freezes out on a thin cylindrical 
shell along which the radial flow shows an azimuthal modulation 
with $|v_x| > |v_y|$. Assuming boost-invariant longitudinal expansion
and freeze-out at constant proper time, the Cooper-Frye freeze-out 
spectrum can be calculated by trivially generalizing the 
derivation given in \cite{R88,HLS90}. In the Boltzmann 
approximation\footnote{For Bose-Einstein or Fermi-Dirac distributions, 
one simply substitutes Eq.~(\ref{spectrum}) by a sum of terms with $T$ 
replaced by $T/n$, $n=1,2,\dots$, and weighted with $(\pm1)^n$ \cite{R88}.} 
one finds, up to irrelevant constants,
 \bea
 \label{spectrum}
     {dN\over dy\,dm_t^2\,d\phi_p} \sim 
     \int_0^{2\pi} \!\!\! d\phi_s\, 
     K_1(\beta_t(\phi_s))\, e^{\alpha_t(\phi_s)\cos(\phi_s{-}\phi_p)} \,,
 \eea  
where $\phi_s,\phi_p$ are the azimuthal angles in coordinate and momentum 
space, and the arguments $\alpha_t(\phi_s){=}$ $(p_t/T)\sinh(\rho(\phi_s))$,
$\beta_t(\phi_s){=}(m_t/T)\cosh(\rho(\phi_s))$ \cite{HLS90} now depend on a 
$\phi_s$-dependent radial flow rapidity $\rho(\phi_s)$. The elliptic flow
coefficient is obtained by taking the azimuthal average over $\cos(2\phi_p)$
with this spectrum, $v_2 = \langle \cos(2\phi_p) \rangle$. The 
$\phi_p$-integral can be done analytically:
 \bea
 \label{v2shell}
   v_2(p_t) = 
   {\int_0^{2\pi} d\phi_s\,\cos(2\phi_s)\,I_2(\alpha_t(\phi_s))
                                        \,K_1(\beta_t(\phi_s))
   \over
   \int_0^{2\pi} d\phi_s\,I_0(\alpha_t(\phi_s))\,K_1(\beta_t(\phi_s))}\,.
\eea 
Making the {\em Ansatz} $\rho=\rho_0 + \rho_a\cos(2\phi_s)$ we checked 
that, for a suitable choice of the average radial flow rapidity $\rho_0$ 
and its azimuthal modulation amplitude $\rho_a$, Eq.~(\ref{v2shell}) 
reproduces all relevant features of the full hydrodynamic results almost 
quantitatively. The $\phi_s$-independent version of Eq.~(\ref{spectrum}) 
\cite{LHS90} has been used very successfully by many groups to fit 
single-particle spectra from central heavy-ion collisions in order to 
extract the average temperature and radial flow velocity at freeze-out. 
Similarly, Eqs.~(\ref{spectrum}) and (\ref{v2shell}) can be used to fit
the spectra and elliptic flow from non-central collisions in order to
extract also the average azimuthal flow modulation. 

\begin{figure}
\begin{center}
 \begin{picture}(200,100)
    \put(65,50){\circle{15}}  \put(65,50){\vector(-1,0){25}} 
                              \put(35,40){$-v_x$}
    \put(135,50){\circle{15}} \put(135,50){\vector(1,0){25}}
                              \put(150,40){$v_x$}
    \put(100,75){\circle{15}} \put(100,75){\vector(0,1){20}}
                              \put(105,85){$v_y$}
    \put(100,25){\circle{15}} \put(100,25){\vector(0,-1){20}}
                              \put(105,10){$-v_y$}
 \end{picture}
\end{center}
  \caption{Simple source of four fireballs.}
  \label{F5}
\end{figure}
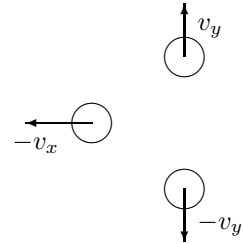

Unfortunately, the remaining angular integral in (\ref{v2shell}) 
cannot be done analytically. To reach a fully analytical understanding 
of many of the features observed in the previous section one can exploit
an even simpler model which still captures the main effects qualitatively,
but no longer quantitatively. It consists of four non-expanding fireballs 
of equal volume, freezing out instantaneously and simultaneously in the
laboratory frame, whose centers move in the transverse plane without 
longitudinal velocity component (see Fig.~\ref{F5}). What matters for the 
observed particle flow pattern are the velocities of the four fireballs, 
as indicated in the figure, but not their positions nor the size of their 
common laboratory frame volume. 

\begin{figure}
\begin{center} 
    \epsfxsize 7.5cm \epsfbox{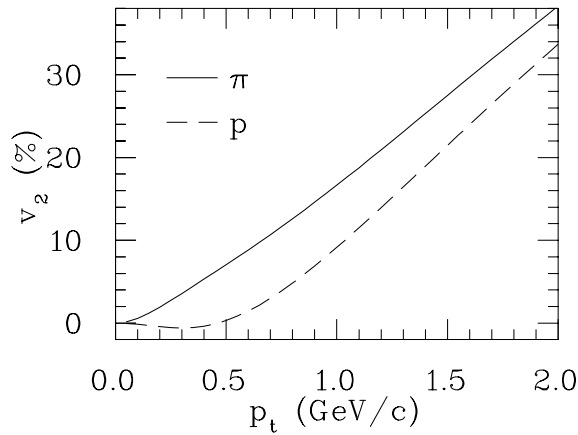}
\end{center}
\caption{Transverse momentum dependence of elliptic flow for 
         midrapidity pions and protons from the schematic source
         in Figure~\protect\ref{F5}, for $T=140$\,MeV, $v_x=0.6$, 
         and $v_y=0.5$.}
\label{F6}
\end{figure}

For this model one finds, again in the Boltzmann approximation, the 
simple expression 
 \begin{eqnarray}
   v_2({\rm y},p_t) & = & 
   \frac{I_2\left(\frac{\gamma_x v_x p_t}{T}\right) 
         - e^{\frac{E}{T}(\gamma_x-\gamma_y)} 
              I_2\left(\frac{\gamma_y v_y p_t}{T}\right)} 
        {I_0\left(\frac{\gamma_x v_x p_t}{T}\right) 
         + e^{\frac{E}{T}(\gamma_x-\gamma_y)}
              I_0\left(\frac{\gamma_y v_y p_t}{T} \right)}.  
   \label{schemav2} 
 \end{eqnarray} 
In this equation the particle mass enters only in the term
$e^{\frac{E}{T}(\gamma_x-\gamma_y)}$, and it is easy to see 
that if all other variables are held fixed, $v_2$ decreases 
with increasing mass. Figure~\ref{F6} shows that our schematic 
source also reproduces the feature of negative $v_2$ for protons
at small $p_t$ which was anticipated above from the ``blast
wave'' model. (This feature is also preserved by the cylindrical 
shell model, Eq.~(\ref{v2shell}).) The ``blast wave peak'' in the 
single-particle spectrum is known to disappear when the constant 
expansion velocity is replaced by a realistic radial velocity 
distribution \cite{LHS90}; we therefore then also expect the dip of 
$v_2$ to negative values to be washed out, in agreement with 
Figs.~\ref{F3} and \ref{F4}. On the other hand, the UrQMD calculations 
reported in \cite{Bleicher00} do show negative $v_2$ for midrapidity 
protons at small $p_t$; it would be interesting to further investigate 
the origin of this effect in that model. It may be worth pointing out
that a similar interplay between radial flow, random thermal motion 
and a {\em directed} flow anisotropy (first instead of second harmonic 
modulation) has been shown \cite{Voloshin97} to be responsible for
negative directed flow coefficients $v_1$ at low $p_t$ \cite{NA49v2}.
 
Our schematic source also permits us to understand the approximately
linear $p_t$-dependence of $v_2$ in the intermediate $p_t$ region. 
Expanding the Bessel functions in (\ref{schemav2}) for large arguments
and keeping only the leading term gives for midrapidity (y=0) particles
  \begin{eqnarray}
   v_2(p_t) & \approx &
      \tanh\left(\frac{1}{2}
                 \left(\frac{\kappa p_t - \lambda m_t}{T} + \mu\right)
           \right),
  \label{linear}
  \end{eqnarray}
where $\kappa = \gamma_x v_x-\gamma_y v_y$, $\lambda = \gamma_x -\gamma_y$ 
and $\mu = \ln(\sqrt{\frac{\gamma_x v_x}{\gamma_y v_y}})$. Since $\kappa
> \lambda$ for $|v_x| > |v_y|$, we get for $p_t \gg m$ the simple form
$v_2(p_t)\approx \tanh(\xi p_t)$ with $\xi = (\kappa-\lambda)/2T$.
For the example in Fig.~\ref{F6}, $\xi \approx 5$\,GeV, and $v_2$ begins 
to turn over at $p_t > 2.5$\,GeV, saturating at $v_2=1$ as $p_t\to\infty$.
This consideration also shows that the scale at which $v_2$ changes from
its required quadratic $p_t$-dependence at low $p_t$ \cite{D95} (with a 
positive or negative coefficient) to a linear behaviour at intermediate 
$p_t$ is given by the particle rest mass $m$, multiplied by a prefactor 
of order 1 which is governed by the average radial flow velocity 
$\la v_\perp \ra$.

We finally discuss the dependence of $v_2$ on the freeze-out temperature.
Taking the temperature derivative of Eq.~(\ref{linear}) we get (again at 
y=0)
  \begin{equation}
   \frac{\partial v_2}{\partial T}(p_t) \approx
   \frac{\lambda m_t - \kappa p_t}{2 T^2} 
    \frac{1}{\cosh^2
    \left(\frac{1}{2}
          \left(\frac{\kappa p_t - \lambda m_t}{T}+\mu\right)
   \right)}\,.
  \label{dTf}
  \end{equation}
With $\kappa > \lambda$ one sees that at fixed $p_t$ the sign of the 
derivative depends on the particle mass and that, for increasing 
$T_{\rm f}$, $v_2(p_t)$ may thus decrease for pions while increasing
for protons, as seen in Fig.~\ref{F4}.\footnote{Of course, in a full 
hydrodynamic calculation $T_{\rm f}$ cannot be varied independently
of the flow velocities, and the latter are not fixed, but vary over
the freeze-out surface. Eq.~(\ref{dTf}) can only give a qualitative 
impression.}

{\it 4. Summary.--}
We have presented a variety of predictions for the elliptic flow and 
single-particle spectra for different hadron species produced in Au+Au 
collisions at $\sqrt{s}=130\,A$\,GeV, using a relativistic hydrodynamic 
model. We have studied the sensitivities to the equation of state and 
freeze-out temperature and showed that these can be used to further
constrain the model parameters and test the approach on a quantitative 
level. A simple expression for fitting spectra and elliptic flow data 
in order to extract the average radial flow and flow anisotropy has been
given. Crucial features of the $p_t$-dependence of the elliptic flow
have been elucidated with a simple schematic model. Testing the 
predicted $p_t$-dependence of $v_2$ for many different hadron species
will clarify the validity of the picture of a thermalized expanding 
source with a common flow velocity for all hadrons at RHIC energies.

{\it Acknowledgments:} Fruitful discussions with M.~Blei\-cher,
H.~Heiselberg, T.~Hirano, V.~Koch, A.~Poskanzer, R.~Snellings, and
N.~Xu are gratefully acknowledged. This work was supported by the 
Director, Office of Science, Office of High Energy and Nuclear Physics, 
Division of Nuclear Physics, and by the Office of Basic Energy
Sciences, Division of Nuclear Sciences, of the U.S. Department of 
Energy under Contract No. DE-AC03-76SF00098.


\end{document}